\documentclass[11pt,twoside]{article}

\usepackage{asp2004}
\usepackage{epsf}
\usepackage{psfig}
\usepackage{lscape}

\begin{document}
\title{Modeling the Structure of Hot Star Disks: a Critical Evaluation of the Viscous Decretion Scenario}   
\author{A. C. Carciofi$^1$, J. E. Bjorkman$^2$, A. S. Miroshnichenko$^{2,3}$, A. M. Magalh\~aes$^1$ and K. S. Bjorkman$^2$} 
\affil{$^1$Instituto de Astronomia, Geof\'isica e Ci\^encias
Atmosf\'ericas, Rua do Mat\~ao 1226, Cidade Universit\'aria, S\~ao
Paulo, SP, 05508--900, Brazil}    
\affil{$^2$Ritter Observatory, Department of Physics and Astronomy,
University of Toledo, Toledo, OH 43606-3390, USA}    
\affil{$^3$Department of Physics and Astronomy, University
of North Carolina at Greensboro, P.O. Box 26170, Greensboro, NC 27402--6170, USA} 

\begin{abstract} 
We present self-consistent solutions for the disk structure of classical Be stars. Our disk model is hydrostatically supported in the vertical direction and the radial structure is governed by viscosity ($\alpha$-disks). We perform three-dimensional non-LTE Monte Carlo simulations to calculate simultaneously both the equilibrium temperature and Hydrogen level populations and to solve self-consistently for the density structure of the disk. We discuss the general properties of the solution for the disk structure and test our model against observations of $\delta$ Scorpii. 
Our results confirm that a viscous decretion disk model is consistent with these observations.
\end{abstract}


\section{Introduction}  

Observational evidence indicates that the circumstellar disks of Be stars are geometrically thin, optically thick disks, whose outflow velocities are small in comparison to their rotation
speeds \citep[see][for a recent review]{por03}.  
Dynamically this requires that the disk be pressure supported
in the vertical direction and centrifugally supported in the radial direction.  For this reason the disk structure is most likely to be that of a Keplerian hydrostatically supported disk.  How material is injected at Keplerian speed
into this disk is still unclear, but after it is injected, the gas will
spread outward (and inward if it is injected at intermediate radii), owing
to viscous effects \citep{lyn74}.

In a recent paper, \citet[][hereafter CB]{car06} described a new non-local thermodynamic equilibrium (NLTE) Monte Carlo (MC) radiation transfer code that may be applied to study differentially moving three-dimensional circumstellar envelopes.  

As an initial application of this code, CB studied the temperature structure of the disks around Be stars. 
They found the disks to be highly non-isothermal, with a difference of up to 3 times between the minimum and maximum temperatures.
The minimum temperatures are about 30\% of the stellar effective temperature ($T_{\rm eff}$) and occur at the midplane typically around 3---5 stellar radii away from the star, depending on the disk density. The maximum temperatures occur at the base of the star and for very dense disks can be even larger than $T_{\rm eff}$ due to the backwarming of the star.

CB found that the temperature is a hybrid between that of optically 
thick disks of Young Stellar Objects and optically thin winds from Hot Stars.  
For the optically thin disk's upper layers (defined as the regions where the distance to the midplane is larger than one scale-height), the temperature is nearly isothermal, with an average temperature of about 50\% of $T_\mathrm{eff}$. This is much cooler than the 80\% value frequently quoted in the literature  \citep[e.g.,][]{wat86}.

At the midplane the temperature shows a complicated structure. 
It initially drops very quickly close to the stellar photosphere, and the temperature profile is well described by a flat blackbody reprocessing disk \citep{ada87},
\begin{equation}
T_{\rm d}(\varpi)  = 
\frac{T_\star}{\pi^{1/4}}
\left[
\sin^{-1}\left( \frac{R_\star}{\varpi} \right) - 
\frac{R_\star}{\varpi}
\sqrt{1-\frac{R_\star^2}{\varpi^2}}
\right]
^{1/4},
\label{als}
\end{equation}
where $T_\star$ is the temperature of the radiation that illuminates the disk. 
When the disk becomes optically thin vertically, the temperature departs from this curve and rises back to the optically thin radiative equilibrium temperature, which is approximately
constant in the winds of Hot Stars.  

The results for the disk temperature raises the question of how it affects the disk structure. 
Determining the self-consistent solution for a Keplerian disk is a 
somewhat complicated problem, because the disk temperature controls its geometry (via the hydrostatic equilibrium equation), which in turn determines the heating and hence the temperature of the disk \citep{ken87}. 

In this paper we present the first results of an updated version of the code described in CB. In this version of the code we self-consistently solve both the disk temperature and the fluid equations to obtain the disk density structure.

In Section 2 we describe the physical properties of our disk model. In Section 3 we briefly describe our MC code. The properties of the solution for a typical Be star disk model are described in Section 4. Finally, in Section 5 we test our model by fitting continuum observations of $\delta$ Scorpii.

\section{Keplerian Disks}

A popular  model for Be star disks is the viscous decretion disk of \citet*{lee91}.  This model is essentially the same as that employed for 
protostellar disks, the primary difference being that Be disks are outflowing, while pre-main-sequence disks are inflowing.  
In this model, it is supposed that some (as yet unknown) mechanism(s) injects material at the Keplerian
orbital speed into the base of the disk.  
Eddy/turbulent viscosity then transports angular momentum outward from the inner boundary of 
the disk (note that this requires a continual injection of angular 
momentum into the base of the disk).  If the radial density gradient of the 
disk is steep enough, angular momentum is added to the individual fluid
elements and they slowly move outward and form the disk.

To critically test such decretion disk models of Be stars against observations, we must determine the structure of the disk, which, as stated in the introduction, is affected by the temperature.

The structure of an isothermal viscous decretion disk was already studied in the literature \citep[see, for example, ][]{bjo97}. What governs the vertical structure of the disk is the gas pressure, responsible for the Gaussian term in the formula for the disk density (written in cylindrical coordinates $\varpi$, $z$ and $\phi$)
\begin{equation}
\rho(\varpi,z) = \frac{\Sigma(\varpi)}{\sqrt{2\pi}H(\varpi)}e^{-0.5(z/H)^2}.
\label{eq:density}
\end{equation}
The disk surface density, $\Sigma$, is given by
\begin{equation}
    \Sigma(\varpi)=\frac{{\dot M} V_{\rm crit} R_{\star}^{1/2}}
           {3 \pi \alpha a^2 \varpi^{3/2}}
           \left(\sqrt{R_0/\varpi}-1\right)  ,
\label{eq:disk_Sigma}
\end{equation}
This expression shows that, for an isothermal disk, the radial structure is controlled by $\alpha$, the viscosity parameter of \citet{ss73} and $R_0$, the disk "truncation" radius. The density scale is controlled primarily by the equatorial mass loss rate $\dot{M}$, but also by the critical velocity of the star, $V_{\rm crit}$, and the disk temperature via the sound speed $a$. For large disks (i.e., large $R_0$), we have that  the surface density of an isothermal disk is a power-law in radius, $\Sigma\propto \varpi^{-2}$.

The disk scale-height, $H$, is given by
\begin{equation}
                H(\varpi)=(a/v_\phi)\varpi  . \label{eq:scale_height}
\end{equation}
Since the disk is isothermal and the $\phi$ component of the velocity is Keplerian ($v_\phi=V_{\rm crit} \varpi^{-1/2}$), we find the familiar result that an isothermal disk flares as $H\propto\varpi^{1.5}$. Hence, it follows from Eq.~\ref{eq:density} that for an isothermal disk the density falls very sharply with radius as $\rho \propto \varpi^{-3.5}$.

To conclude this brief description of the properties of isothermal Keplerian viscous disks, we must describe their velocity structure. The $\phi$ component is Keplerian, as stated above, but those disks must have an outflow velocity. This velocity is given by the mass conservation relation
\begin{equation}
            v_\varpi = \frac{\dot M}{2 \pi \varpi \Sigma}. 
            \label{eq:radvel}
\end{equation}

As mentioned in the introduction, CB found that the disks of classical Be stars are highly non-isothermal. A temperature structure affects the density structure of the disk via the sound speed in Eqs.~\ref{eq:disk_Sigma} and~\ref{eq:scale_height}.
It is straightforward to show that the formulae for non-isothermal viscous disks are the same as Eqs.~(\ref{eq:density}) to~(\ref{eq:radvel}), with the temperature replaced by the weighted average of $T$ in the vertical direction \citep[see][for details]{car06b}
\begin{equation}
	\langle T \rangle_\varpi =
	 \frac{\int_{-\infty}^{\infty} T(\varpi,z)\rho(\varpi,z) dz}
	{\int_{-\infty}^{\infty} \rho(\varpi,z) dz}.
	\label{eq:avT}
\end{equation}


\section{Monte Carlo Code}

For the results shown in sections 4 and 5 we employ an updated version of the code of CB.
The basic steps of the code are as follows:

\begin{enumerate}
\item The code starts with a constant temperature for the disk, an initial guess for the Hydrogen level populations and the isothermal viscous decretion density [Eqs.~(\ref{eq:density}) to~(\ref{eq:radvel})];
\item A full NLTE MC radiation transfer, using the Sobolev approximation for the line opacities, is performed. 
\item The heating, ionization and excitation rates sampled during step 2 are used to solve the rate equations and the energy conservation equation, to obtain news values for  the disk temperature and Hydrogen level populations.
\item The new disk temperature is used for calculating the new density structure.
\item Steps 2 --- 4 are repeated until all the state variables and the disk temperature have converged to a solution.
\item If wanted, a post-processing code is run to calculate observables (SED, line profiles, images, etc.) using the results of step 5.
\end{enumerate}

\section{Properties of the Solution}

\begin{table}[!ht]
\caption{Parameters of the disk and the central star}
\smallskip
\begin{center}
{\small
\begin{tabular}{ccccccc}
\noalign{\smallskip}
\tableline
\noalign{\smallskip}
$T_{\rm eff}\;(\rm K)$ & 
$R_{\star}\;(R_{\sun})$ &
$\dot{M}\;(M_{\sun}/\rm year)$ &
$\alpha$ &
$V_{\rm crit}\;(\rm km/s)$ &
$R_0\;(R_{\star})$ &
$\rho_0\;(\rm g\;cm^{-3})$ \\

$19,000$ &
$5.6$ & 
$5\times10^{-12}$& 
0.1& 
$560$ & 
$1000$ &
$3.7\times 10^{11}$\\
\noalign{\smallskip}
\tableline
\end{tabular}
}
\end{center}
\label{tab:disk_model}
\end{table}

\begin{figure}[!ht]
\plotone{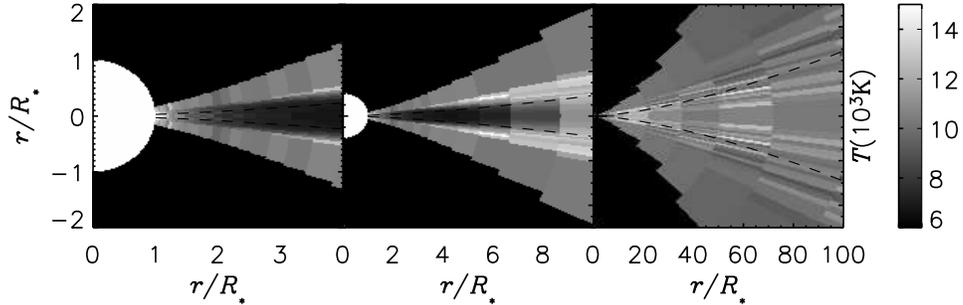}
\caption{Temperature distribution. Each panel shows the temperature as a function of $x$ and $z$ for a different radial scale. 
}
 \label{fig:temp}
\end{figure}

\begin{figure}[!ht]
\plottwo{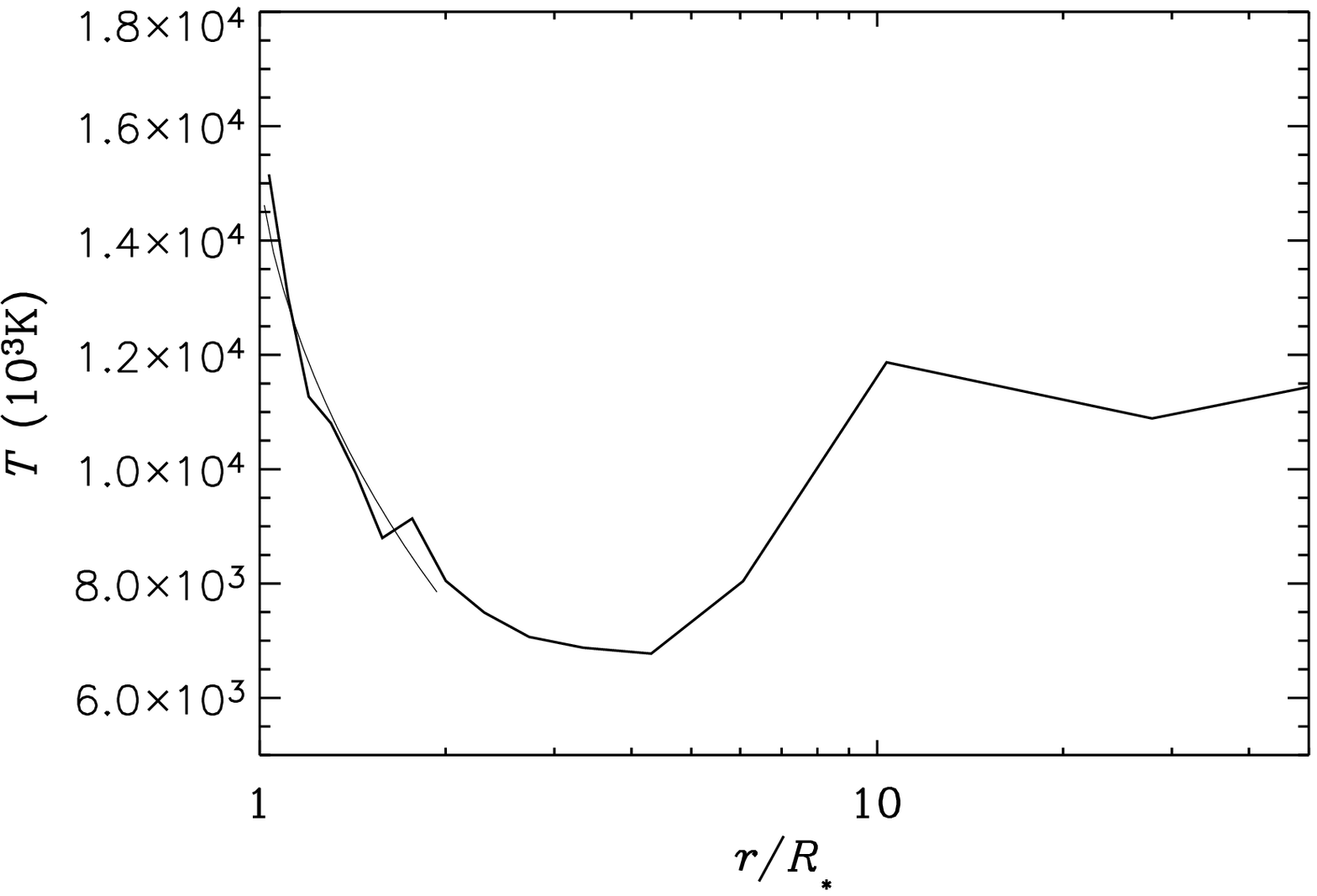}{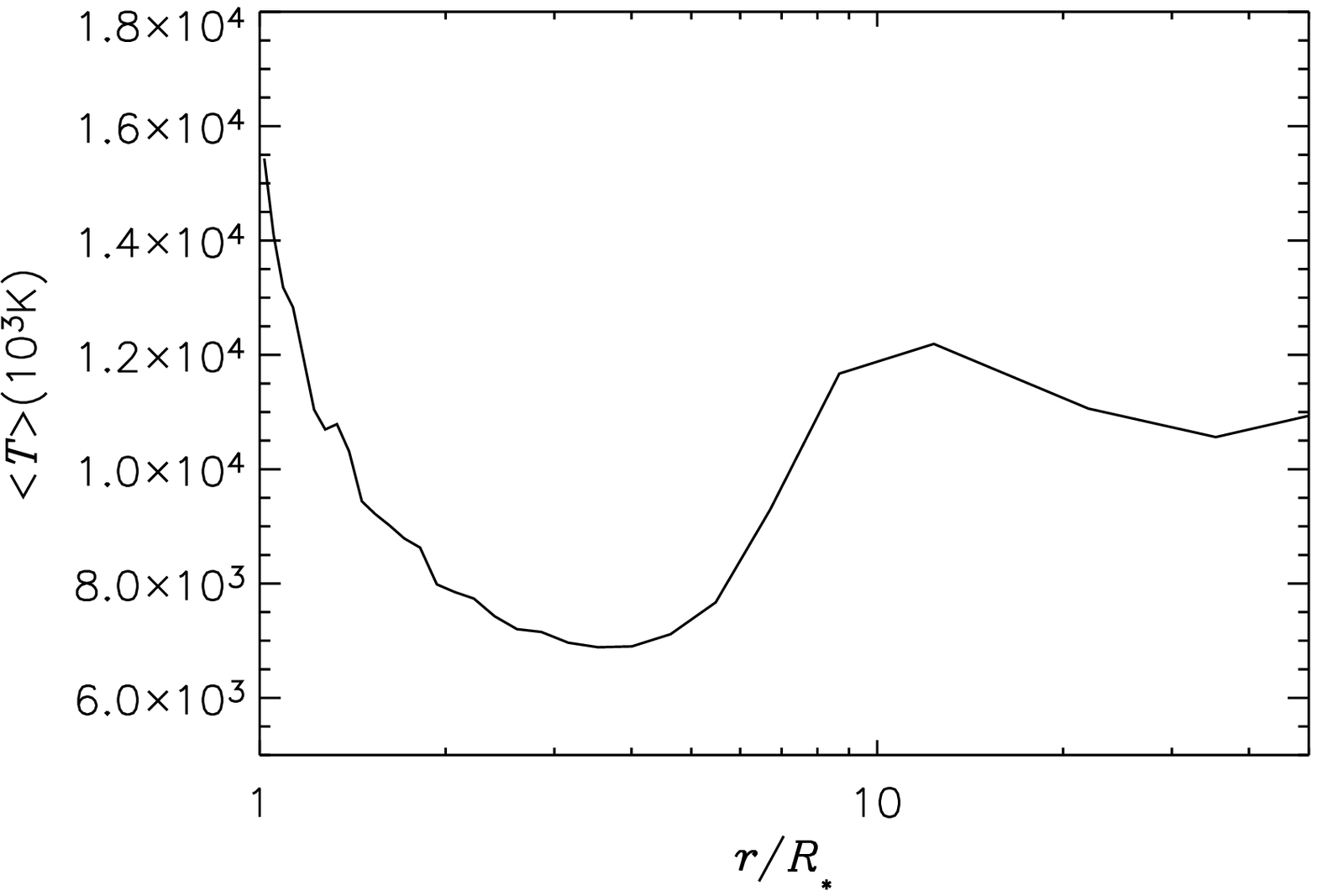}
\caption{Radial temperature structure. {\it Left:} Temperature along the midplane. {\it Right:} Weighted-average of T in the vertical direction [Eq.~(\ref{eq:avT})]. }
 \label{fig:temp2}
\end{figure}

In table~\ref{tab:disk_model} we list the parameters of the disk model we choose as an example to illustrate the properties of our solution for the disk structure. The model is similar to the ones used by CB and consists of a B3\,IV star surrounded by a steady state disk, sustained by an equatorial mass loss rate of $5\times10^{-12} \;M_{\sun}/\rm year$. 
We also carried out calculations for other spectral types and mass loss rates. The details of the solution varies with the choice of parameters, but the general properties of the disk structure are similar. 

CB studied the temperature properties of disks with a known density function, corresponding to the isothermal viscous disk of Eqs.~(\ref{eq:density}) to (\ref{eq:radvel}). 
The temperature structure of our self-consistent Keplerian disk is shown in Figures~\ref{fig:temp} and~\ref{fig:temp2}. 
Interestingly, the temperature structure is nearly identical to the results of CB (see, for instance, their Figures 6 and 7). 
The disk is isothermal in the upper layers and also in the vertically optically thin portions of the midplane ($\varpi/R_{\star}>10$). 
The inner disk ($\varpi/R_{\star}<10$) behaves initially as a flat reprocessing disk [eq.~(\ref{als}), thin line of Figure~\ref{fig:temp2}-a] and departs form this curve when it starts to become optically thin vertically, which happens for $\varpi/R_{\star}\approx3$ (see CB for further details). 

\begin{figure}[!ht]
\plottwo{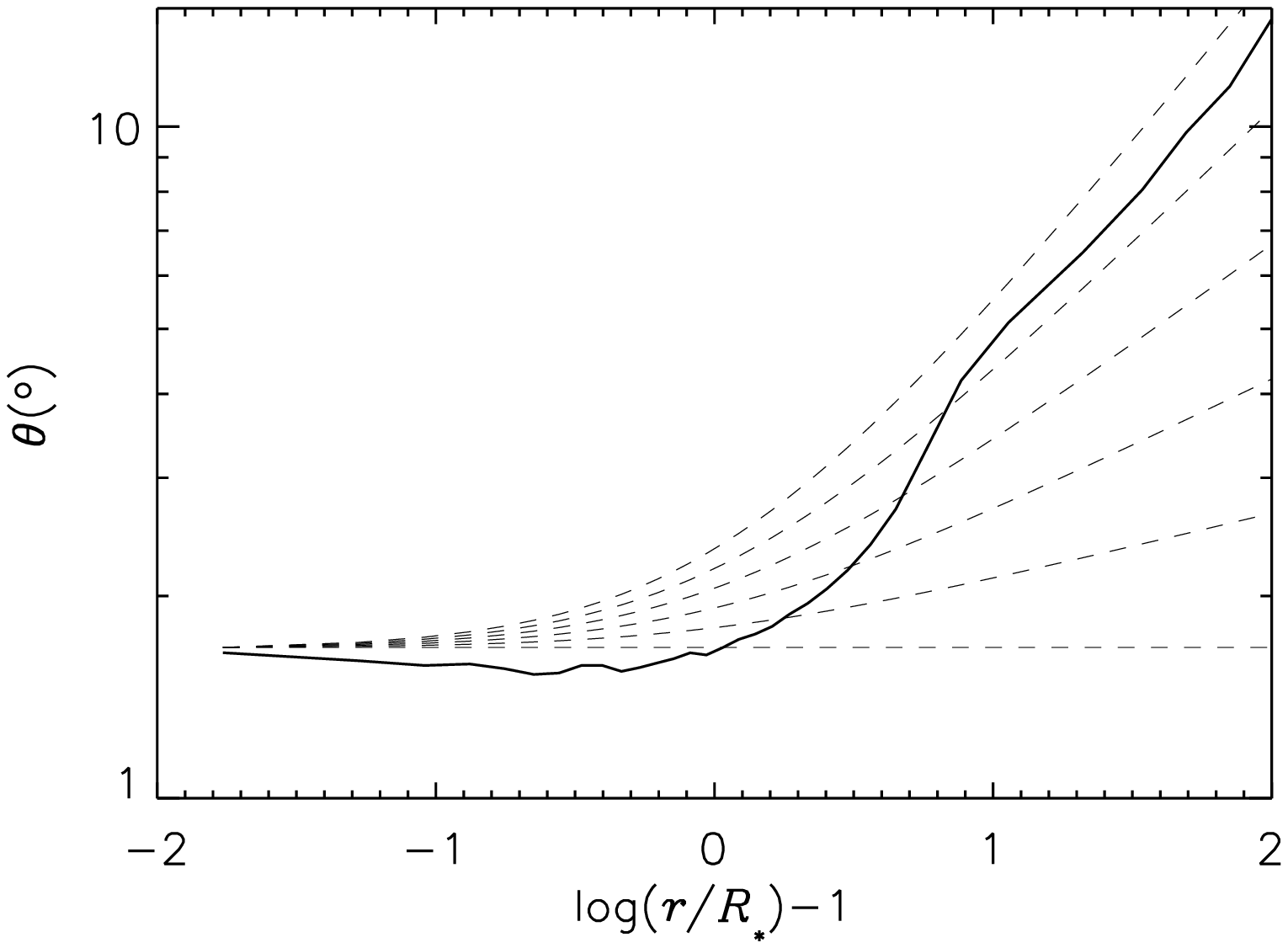}{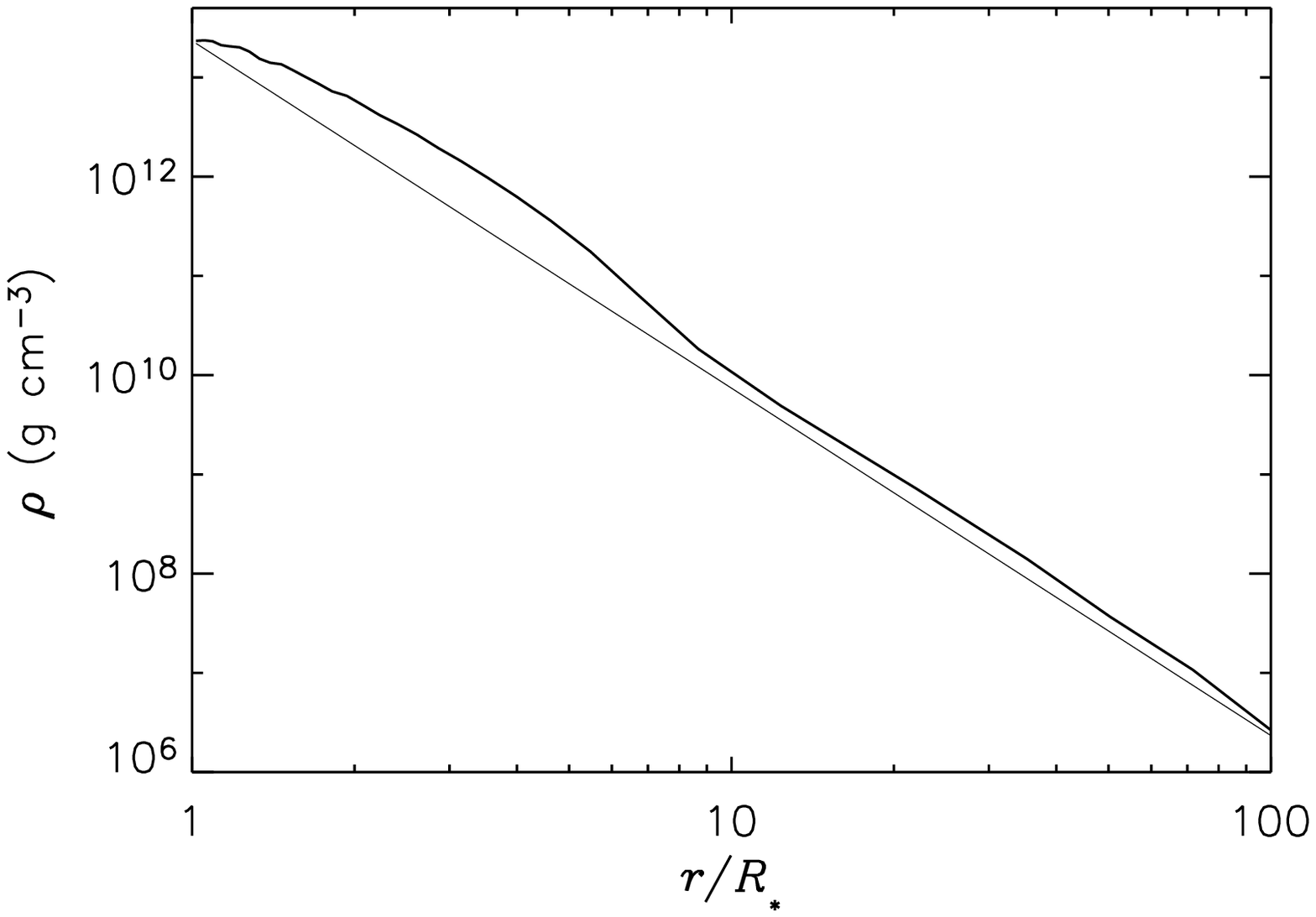}
\caption{Disk density structure. 
{\it Left:} Opening angle. The thick line corresponds to the self-consistent solution for the disk opening angle [eq.~(\ref{eq:opangle})]. The dashed lines correspond to opening angles for scale-heights parameterized as $H(\varpi) \propto \varpi^\beta$. From bottom to top, $\beta=1.0$ to 1.5, in steps of 0.1. 
{\it Right:} Density profile. We compare the density calculated from our MC simulations (thick line) with the $\rho\propto\varpi^{-3.5}$ isothermal density (thin line).
}
 \label{fig:angle}
\end{figure}

In Figure~\ref{fig:angle} we analyze the properties of our solution for the disk density to understand how our self-consistent solution differs from the isothermal solution. 
In the left panel we show the disk opening angle vs. radius, defined as
\begin{equation}
\theta(\varpi) = \sin^{-1}\left[\frac{H(\varpi)}{\varpi}\right]. \label{eq:opangle}
\end{equation}
For reference, we also show the opening angle corresponding to disk scale-heights given by $H(\varpi)\propto\varpi^\beta$. A $\beta$ of 1 corresponds to a non-flared disk (wedge-shaped) and a $\beta$ of 1.5 to an isothermal disk. 

The opening angle for the inner disk ($\varpi/R_{\star} < 10$) is very small, in agreement with the general belief that the disks of Be stars are geometrically very thin. 
However, the behavior of the opening angle is rather peculiar; we observe a \emph{decrease} of the opening angle (de-flaring) between 1 and 2 stellar radii. This de-flaring is a consequence of the very sharp fall of the temperature (see Figure~\ref{fig:temp2}).
The disk flares significantly only for $\varpi/R_{\star} > 10$, where it becomes isothermal.

The radial structure of the disk density is shown in the left panel of Figure~\ref{fig:angle}. We also show for comparison the $\rho \propto \varpi^{-3.5}$ curve corresponding to the isothermal solution for the density. We see that the slope of the MC density profile for the inner disk departs significantly from the -3.5 value; it is much shallower close to the star and much steeper for $4 < \varpi/R_{\star}<10$. 
This is, again, a result of the temperature structure in the midplane. 
From the observational point of view this is an interesting result, because the slope of the density profile can, in principe, be mapped via a detailed study of the IR excess \citep[][]{wat86}.

\section{The Case of $\delta$ Scorpii \label{deltaSco}}

\begin{figure}
\plottwo{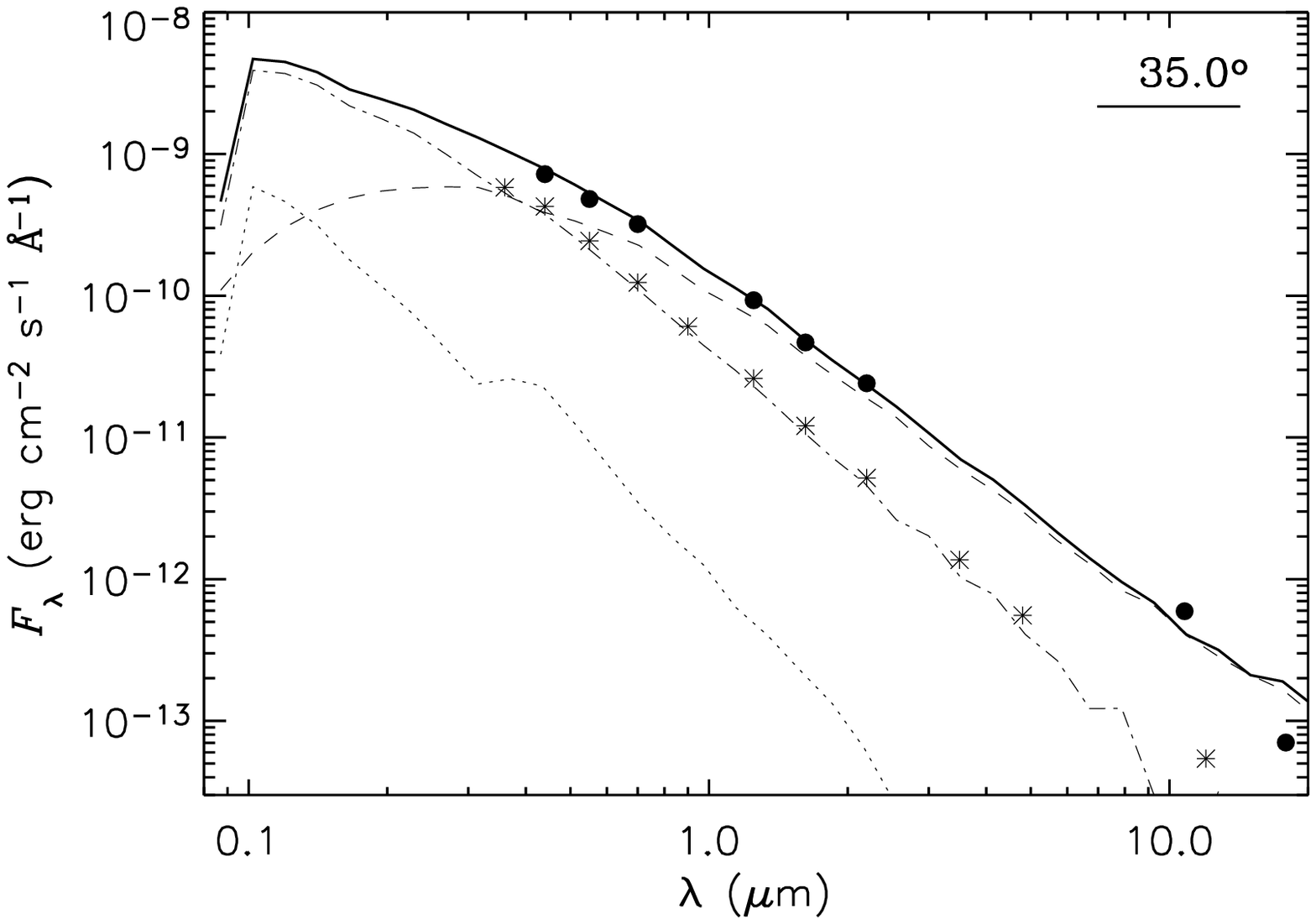}{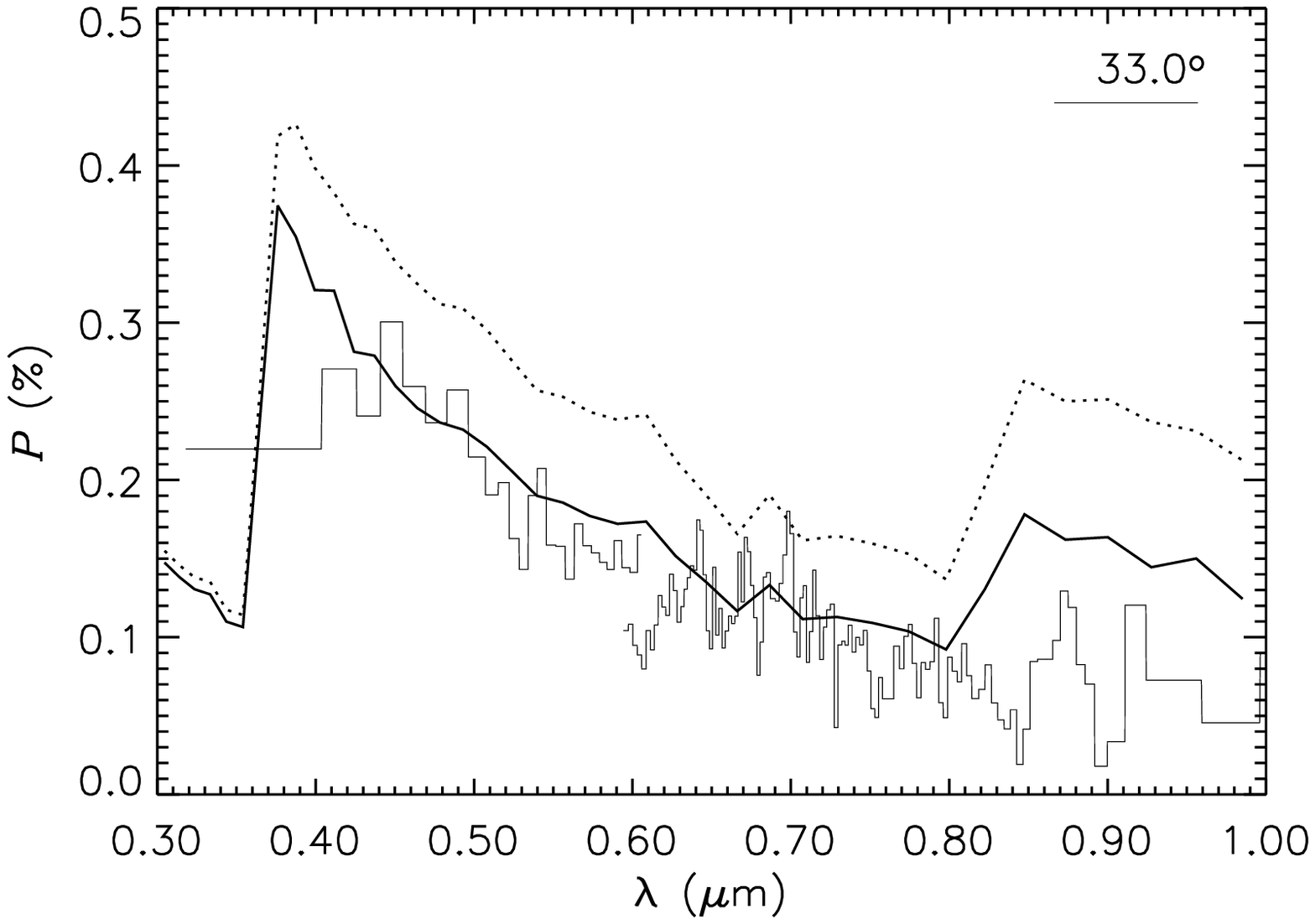}
\caption{Best-fitting model. 
{\it Left:} 
Our best-fitting SED (solid line) is shown along with
the observed active-phase SED (filled circles) and the pre-active phase SED (stars).
The other lines  correspond to the scattered, emitted and attenuated stellar fluxes
(dotted, dashed, and dash-dot lines, respectively).
{\it Right:}
The best-fitting model polarization (thick lines) and the
observed polarization (thin line) are shown. The solid thick line correspond to the model
polarization with the depolarizing effect of the secondary included, and the dotted thick
line correspond to the uncorrected polarization.
}
 \label{fig:dsco_sed}
\end{figure}

\begin{table}[!ht]
\caption[]{Stellar Parameters for $\delta$ Scorpii} 
\begin{center}
{\small
\begin{tabular}{ccccc}
\tableline
\noalign{\smallskip}
$R_{\star}\;(R_{\sun})$ & $T_\mathrm{eff}\;\rm K$ & $M_{\star}\;(M_{\sun})$ & $V_\mathrm{crit}\;$(km/s) & $i\;(^{\circ})$ \\
$7$ & 27,000 & $14$ & 620 & $38\pm5$ \\ 
\noalign{\smallskip}
\tableline
\end{tabular}
}
\end{center}
\label{tab:dsco}
\end{table}

\begin{table}[!ht]
\caption[]{Best-fit Disk Parameters for $\delta$ Scorpii}
\begin{center}
{\small
\begin{tabular}{cccc}
\tableline
\noalign{\smallskip}
$\dot{M}\;(M_{\sun}/\mathrm{year})$ & $\rho_0\;(\mathrm{g}/\mathrm{cm}^3)$ &  $R_\mathrm{d}\;(R_\star)$ & $i\;(^{\circ})$ \\
$2.5 \times 10^{-9}$ &
$5.2\times 10^{-10}$ &
$5$ &
 $33$ \\
\noalign{\smallskip}
\tableline
\end{tabular}
}
\end{center}
\label{tab:dsco2}
\end{table}

To critically test our model against observations, we model continuum observations of $\delta$ Scorpii. 
This star has been a standard of spectral classification (B0
{\sc IV}) since long ago, but in the summer of 2000 it started to show signs of circumstellar activity that have lasted to present day. The object is a non-eclipsing binary system with a $\sim$1.5 mag optically fainter secondary companion with an orbital period of 10.6 years
\citep{bed93} and a highly eccentric orbit \citep[$e$=0.94,][]{mir01}.
We use in our modeling data corresponding to the SED before and after the Be phase.
We also model the continuum polarization between $3000$ and $10000\;$\AA\,
 \citep[see][for further details]{car06c}.

The parameters of the primary companion were extracted from \citet{mir01}
and are summarized in Table~\ref{tab:dsco}.
We assume that the disk around the primary companion is in the orbital plane of the binary system,
so the disk inclination angle is the same as the orbital inclination angle $i$. 

Figure~\ref{fig:dsco_sed} shows our best-fitting SED and polarization along with the observed data. 
The parameters of our best-fitting model are listed in Table~\ref{tab:dsco2}.
Our model reproduces the flux observations quite well. A particularly significant result is the fact that the change in slope between the visible and the NIR fluxes is well described by the model.
Also, the model reproduces well the polarization in the Paschen continuum (left curve of Figure~\ref{fig:dsco_sed}).

In \citet{car06c} we discuss the above results in detail. There are some shortcomings in our modeling that we do not discuss here, given the limited scope of this paper. However, the fact that our models reproduce well the observations for $\delta$ Scorpii is an important evidence that this star (and probably the other Be stars) is surrounded by a non-isothermal viscous Keplerian disk.

\acknowledgements 
ACC and AMM acknowledges support from the FAPESP (grants 01/12589-1 and 04/07707-3). AMM acknowledges support from CNPq.
JEB acknowledges support from NSF grants AST-9819928 and AST-0307686.
ASM and KSB acknowledge support from NASA grant NAG5--8054.


\end{document}